\documentclass{article}
\usepackage[utf8]{inputenc}

\usepackage{amsmath}
\usepackage{amssymb}
\usepackage{physics}
\usepackage{multirow}
\usepackage{fancyhdr}

\newcommand{\etal}{\textit{et al}.~}
\newcommand\numberthis{\addtocounter{equation}{1}\tag{\theequation}}
\newcommand{\horrule}[1]{\rule{\linewidth}{#1}}

\title{Quantumizing Classical Games:\\ An Introduction to Quantum Game Theory}
\author{}
\date{}

\begin{document}

\begin{titlepage}

\begin{center}
 \Large
% {\color{RawSienna}
% {\textsf{Imperial College London}}}\\
% \normalsize
% \large
% {\textsc{Dept. of Electrical and Electronic Enginnering}\\}

% \vspace{0.3cm}
% \vspace{.1in}

% \textmd{ EEE 306}\\
% \textmd{{ Power Systems I Laboratory} }\\[0.3in]

\horrule{0.5pt} \\[0.5cm] % Thin top horizontal rule
{\Large \textsf{Quantumizing Classical Games:}}\\ 
{\textsf{An Introduction to Quantum Game Theory}}\\% The assignment title
\horrule{1.5pt} \\[0.5cm] % Thick bottom horizontal ru
% Title
%\Large \textbf {\textsc{Power Plant Visit Report}}\\[0.5in]
\vspace{.5in}
\large

% \textsf{Submitted by}\\[0.1in]
% {\large\textsc{Sowmitra Das}}\\
\textsf{Sowmitra Das}\\
\textsf{MSc, Applied Mathematics}\\
\textsf{Imperial College London}\\[0.2in]

\textsf{Email: sowmitra.das22@imperial.ac.uk}\\[0.2in]
\vfill
\vspace{.2in}
\textsf{As submission of Course-Project for :}\\[0.1in]
\textsf{MATH70007: Dynamics of Learning \& Iterated Games}\\
\textsf{Autumn 2022}
\vspace{1in}
\end{center}

\end{titlepage}

% \maketitle

\tableofcontents

\newpage

\begin{abstract}
\normalsize
    We give a concise and self-contained introduction to the theory of Quantum Games by reviewing the seminal works \cite{meyer1999quantum, eisert2000quantum, marinatto2000quantum, landsburg2004quantum} which initiated the study of this field. By generalizing this body of work, we formulate a protocol to \emph{Quantumize} any finite classical $n$-player game, and use a novel approach of describing such a Quantum Game in terms of commuting Payoff Operators. We describe what advantages can be gained by players by quantumizing such a game, particularly, what additional Nash Equilibria the players can achieve and the Pareto-Optimality of these additional equilibria. 
\end{abstract}

\section{Introduction}
Quantum Information Theory is the study of Information Processing on Quantum Systems, utilizing the effects of Quantum Mechanics. This encompasses novel domains such as Quantum Computation \cite{nielsen_chuang_2010}, Quantum Communication \cite{bennett1984g, gisin2007quantum}, Quantum Metrology \cite{simon2017quantum}, etc. Particularly, the principles of of Quantum Information could be applied to any domain which uses computation, communication or information-processing of any sort to perform a specific task. This brings us to the study of Game Theory and its extension to Quantum Games. 

Game Theory concerns itself with the strategic behaviour of rational agents who are interacting with each other to achieve some payoff from a finite pool of resources. The study of Quantum Games is a reasonable successor to this theory, if the agents have to manipulate quantum systems or have access to quantum effects while deciding on their strategies. such effects include Superposition, Entanglement and/or the randomness of Quantum Measurements. This analysis becomes particularly worthy of interest if these effects and correlations result in strategies that cannot be emulated by classical means, and if these \emph{quantum} strategies result in better payoff for the players. 

We embark on our study of Quantum Games because over the past 2 decades, researchers have shown just that. In 1999, in a seminal paper \cite{meyer1999quantum} Meyer  showed that, it is possible to construct a 2-player Quantum Game, where, if one player is restricted to classical strategies and the other has access to Quantum strategies, the Quantum player always wins. Then in 2000, Eisert, Wilkens and Lewenstein \cite{eisert1999quantum, eisert2000quantum} quantumized another 2-player game, in which they showed that if both players have access to a restricted set of Quantum Strategies, they can achieve a Nash Equilibrium that is Pareto-Optimal to the Classical Nash Equilibrium. Dahl 
\cite{dahl2011quantum}, and later Landsburg \cite{landsburg2004quantum, landsburg2011nash} generalized this to arbitrary 2-player games, and show that the payoff resulting from both players using Quantum Strategies is always at least as great as that obtained from using Classical Strategies, and one can construct scenarios in which it is greater. 

In this expositional article, we review all of these results - by first laying down the principles of Quantum Mechanics and Classical Game Theory (in a form that is suitable for our purposes of generalization), and then provide a framework to combine these two. In this journey, we quantumize the Penny-Flip, Prisoner's Dilemma and the Battle of Sexes games. At the end, we generalize this body of work, and provide a protocol to quantumize any finite $n$-player game. We also introduce a slightly novel device of \emph{Payoff Operators} to describe the $n$-player Quantum Games. We hope this short introduction will be helpful to anyone hoping to enter into research in this field.

\section{Review of Quantum Mechanics}
First, we review the basic principles of Quantum Mechanics and Quantum Information Theory in a form that is suitable for the formulation of Quantum Games. In what follows, we give a concise description of the postulates used in the Quantum-Mechanical description of natural systems.  
\begin{enumerate}
    \item\emph{Postulate 1: States}\\ 
    The fundamental object of interest of a system in Quantum Mechanics is its ``state". The first postulate asserts that, the quantum state of an \emph{isolated} system is a (generally infinite-dimensional) vector in a complex Hilbert space $\mathcal{H}(\mathbb{C})$. However, for our purposes, we will be solely focusing on finite dimensional systems. Of particular interest will be systems of dimension 2, called qubits - aptly named because they are the quantum analogue of their classical counterpart ``bits". So, in Dirac's famous bra-ket notation, the state of a qubit $\ket{\psi}$ is represented as - 
    \[\ket{\psi} = \begin{bmatrix}\alpha \\ \beta \end{bmatrix}\]
    where, $\alpha$ and $\beta$ are complex numbers. However, $\alpha$ and $\beta$ cannot be arbitrary, and have to obey certain constraints which will be outlined in subsequent postulates. 
    
    Particular states of qubits are given special names. The canonical vectors $\begin{bmatrix} 1 \\ 0
    \end{bmatrix}$ and $\begin{bmatrix} 0 \\ 1
    \end{bmatrix}$ are called the $\ket{0}$ and $\ket{1}$ states respectively, thus named because they are generally used to reflect the ``0'' and ``1'' states of a classical bit. However, as is evident, a general quantum state of a qubit can be a complex linear combination of these two states, i.e, in general - 
    \[\ket{\psi} = \alpha\ket{0} + \beta\ket{1}\]
    This is called a ``\emph{superposition}'' state, which is one of the defining features of a quantum system, and which has no classical analogue. 

    \item \emph{Postulate 2: Measurement}\\
    A measurement of an observable $O$ of a Quantum System, \[\mathcal{M}_o = \{(o_1, \Pi_1), (o_2, \Pi_2), \ldots, (o_n, \Pi_n)\}\] 
    is described by a set of outcomes $\{o_1, o_2, \ldots o_n\}$ and a corresponding complete set of orthonormal projection operators $\{\Pi_1, \Pi_2, \ldots, \Pi_n\}$, such that, 
    \[\Pi_i\Pi_j = \delta_{ij}\Pi_i \qquad \text{and} \qquad \sum_i\Pi_i = \mathbb{I}\]
    where, $\delta_{ij}$ is the Kronecker Delta and $\mathbb{I}$ is the Identity Operator. 
    
    One of the most intriguing (and mysterious) features of the Quantum-Mechanical description of nature is that the measurement postulate asserts, the outcome of a Quantum Measurement is not deterministic, rather probalistic. If a measurement of $O$ is performed on a quantum system in a state $\ket{\psi}$, and the measurement outcome $o_i$ has a corresponding projection operator $\Pi_i$, then the \emph{probability} that the outcome $o_i$ will be obtained upon measurement is given by -    
    \[p(o_i) = \bra{\psi}\Pi_i\ket{\psi}\]
    Here, $\bra{\psi}$ is the Hermitian Conjugate of the vector $\ket{\psi}$, i.e, 
    $$\ket{\psi}^{\dagger} = (\ket{\psi}^{\textsf T})^* = \bra{\psi}$$ 

    Due to this probabilistic description, a quantum system (in general) does not have a specific \emph{a priori} value for an observable $O$, and any of the outcomes $o_i$ can be obtained as a measurement result. 

    Another consequence of this postulate is that it provides a constraint on the form of $\ket{\psi}$. Since the rules of probability dictate that, the probabilities of all outcomes must sum to 1, we obtain - 
    \[1 = \sum_ip(o_i) = \sum_i \bra{\psi}\Pi_i\ket{\psi} = \bra{\psi}\left(\sum_i\Pi_i\right)\ket{\psi} = \bra{\psi}\mathbb{I}\ket{\psi} = \bra{\psi}\ket{\psi}\]
    i.e, $\bra{\psi}\ket{\psi} = 1$. This implies that, $\ket{\psi}$ is a normalised vector. Particularly, if we have a qubit state $\ket{\psi} = \begin{bmatrix}\alpha \\ \beta \end{bmatrix}$, then, 
    \[\abs{\alpha}^2 + \abs{\beta}^2 = 1\]

    The second part of the measurement postulate dictates how the state of the system changes under measurement. Similar to the outcome of the measurement, this is also random, and it depends on the outcome. If an outcome of $o_i$ is obtained from a measurement of $O$ on a quantum system in a state $\ket{\psi}$, then the $\emph{post-measurement state}$ \underline{given} $o_i$ is -
    \[\ket{\psi_i} = \frac{\Pi_i\ket{\psi}}{\sqrt{p(o_i)}}\]
    i.e, upon measurement, the state $\ket{\psi}$ instantaneously (and randomly) ``collapses'' to the state $\ket{\psi_i}$ with probability $p(o_i)$. This is famously termed as ``the collapse of the wave-function''. 
    
    At the risk of repitition, we reiterate that, this measurement procedure is ``intrinsically" random and does not represent any ignorance or faulty equipment on the side of the experimenter. In a Quantum-Mechanical description of nature, there is no dynamical equation that can deterministically predict the state to which a collapse will occur. We can only measure probabilities. (Although, there is an evident caveat. Clearly, if the probability of a certain outcome is \emph{1}, we \emph{can} predict with full confidence the state to which the collapse will occur after the measurement.) 

    We demonstrate the measurement postulate with the following example. Suppose, a qubit is in the state - 
    \[\ket{\psi} = \alpha\ket{0} + \beta\ket{1}\]
    We perform a measurement on this state with outcomes and projection operators - 
    \[\mathcal{M} = \{(0, \dyad{0}), (1, \dyad{1})\}\]
    Then, the probabilities of getting the outcomes 0 and 1 are, respectively, 
    \[p(0) = \bra{\psi}\big(\dyad{0}\big)\ket{\psi} = \abs{\bra{0}\ket{\psi}}^2 = \abs{\alpha}^2\]
    \[p(1) = \bra{\psi}\big(\dyad{1}\big)\ket{\psi} = \abs{\bra{1}\ket{\psi}}^2 = \abs{\beta}^2\]
    If the measurement outcome is $0$, the post measurement state is - 
    \[\ket{\psi_0} = \frac{(\dyad{0})\ket{\psi}}{\sqrt{p(0)}} = \frac{\alpha \ket{0}}{\abs{\alpha}} \sim \ket{0}\]
    up to a global phase factor $e^{i\theta}$ which is not important in Quantum Mechanics. Similarly, we can show that, 
    \[\ket{\psi_1} = \ket{1}\]
    This type of measurement using the projectors $\dyad{0}$ and $\dyad{1}$ is also called a ``measurement in the $\{\ket{0}, \ket{1}\}$ basis''. In short, if a state $\ket{\psi} = \alpha\ket{0} + \beta\ket{1}$ is measured in the $\{\ket{0}, \ket{1}\}$ basis, then, $\ket{\psi}$ collapses to $\ket{0}$ with probability $\abs{\alpha}^2$ and it collapses to $\ket{1}$ with probability $\abs{\beta}^2$. Here, $\alpha$ and $\beta$ are called the \emph{probability amplitudes} corresponding to $\ket{0}$ and $\ket{1}$ respectively. 

    \item \emph{Postulate 3: Evolution}\\
    The third postulate describes how a state evolves with time. It dictates that, the evolution of the state of an \emph{isolated} quantum system (i.e, one which is not interacting with the outside world, or, one which is not being measured) is described by a ``linear'' operator, given that normalization is preserved. So, if a state $\ket{\psi}$ evolves into a state $\ket{\psi'}$, then, 
    \[\ket{\psi'} = U\ket{\psi}\]
    for some linear operator $U$. In addition, if $\ket{\psi}$ is normalized, then $\ket{\psi'}$ also has to be normalized. So, we have, 
    \[1 = \bra{\psi'}\ket{\psi'} = \bra{\psi}U^{\dagger}U\ket{\psi}\]
    Since, this needs to be true for all normalized vectors $\ket{\psi}$, we have, 
    \[U^{\dagger}U = \mathbb{I}\]
    Operators which satisfy this condition are called \emph{Unitary} Operators. So, in short, time evolution of an isolated quantum system is described by a Unitary Operator. 

    \item \emph{Postulate 4: Composite Systems}\\
    So far, we have considered the states of systems which are isolated. This postulate provides a means to combine systems, or from another perspective, how to describe systems which are interacting with each other (but are combinedly isolated from the rest of the universe).

    Suppose, a system $A$ is interacting with a system $B$ (the combined system $AB$ being isolated). If, the state of the isolated system $A$ is described by the Hilbert Space $\mathcal{H}_A$ and the state of the isolated system $B$ is described by the Hilbert Space $\mathcal{H}_B$, then the state of the combined interacting system $AB$ is described by the Tensor-Product Hilbert Space $\mathcal{H}_A\otimes \mathcal{H}_B$, where, 
    \[\mathcal{H}_A\otimes \mathcal{H}_B = \text{span}\left\{\ket{\psi}_A\otimes \ket{\phi}_B \Big| \ket{\psi}_A \in \mathcal{H}_A, \ket{\psi}_B \in \mathcal{H}_B\right\}\]
    i.e, $\mathcal{H}_A\otimes \mathcal{H}_B$ consists of linear combinations of normalized tensor-product states of the form $\ket{\psi}_A\otimes \ket{\phi}_B$ over the field of complex numbers $\mathbb{C}$. 

    There are 2 classes of states in the space $\mathcal{H}_A\otimes \mathcal{H}_B$. States of the form $\ket{\psi}_A\otimes \ket{\phi}_B$ are called ``Product States'', and they describe a situation in which the system $A$ is in the state $\ket{\psi}_A$ and the system $B$ is in the state $\ket{\phi}_B$. However, infinitely more interesting, is a second class of states which cannot be factorized into seperate states for the individual systems. A simple example is the following 2-qubit state - 
    \[\ket{\Phi}_{AB} = \frac{1}{\sqrt{2}}\left(\ket{0}_A\otimes \ket{0}_B + \ket{1}_A\otimes \ket{1}_B\right)\]
    As can be checked easily, the state $\ket{\Phi}_{AB}$ cannot be decomposed into any state of the form $\ket{\psi}_A\otimes \ket{\phi}_B$. These type of states are called ``\emph{Entangled}'' states and they represent a scenario in which the combined system $AB$ has a state, but the individual systems do not have a state of their own. Entangled states form the basis of the essential ``quantumness'' in Quantum Information Theory, and all of the recent breakthroughs in Quantum Computing. As we will soon see, they are also at the heart of the advantages of Quantum Strategies over Classical Strategies in game theoretic settings. 
\end{enumerate}

\section{Review of Game Theory}
Before moving on to analyzing the advantages of Quantum Strategies, we review the basic notions of Game Theory and solution concepts in a form that is suitable for generalization to the Quantum Domain. 

A Pure-Strategy $n$-player ``Game'' \textbf{G}$(\mathcal{S}, \pi)$ is described by a tuple of ``strategy" spaces -
\begin{equation}\label{pG_S}
    \mathcal{S} = (\mathcal{S}_1, \mathcal{S}_2, \ldots, \mathcal{S}_n)
\end{equation}
and a tuple of payoff functions - 
% \[\mathcal{P} = (\mathcal{P}_1, \mathcal{P}_2, \ldots, \mathcal{P}_n)\]
\begin{equation}\label{pG_pay}
    \pi = (\pi_1, \pi_2, \ldots, \pi_n)
\end{equation}
Each of the sets $\mathcal{S}_i$ represents the collection of `strategies' player $i$ has access to in the game. This can be the set of moves that player $i$ can play, or the decisions that he/she might take to influence the outcome of the game. The set $\mathcal{S}_i$ can be written as follows -
\begin{equation}\label{pG_strat}
    \mathcal{S}_i = \{s_i^1, s_i^2, \ldots, s_i^{\alpha}, \ldots\}
\end{equation}
where, the $s_i^{\alpha}$ are the pure strategies that player $i$ has access to. Henceforth, we consider ``finite'' games, i.e, games in which each of the sets $\mathcal{S}_i$ are finite. 
We define a ``\emph{play}'' $\mathcal{P}$ of the Game \textbf{G} as an $n$-tuple of strategies 
\[\mathcal{P}(\mathbf{\alpha}):= (s_1^{\alpha_1}, s_2^{\alpha_2}, \ldots, s_n^{\alpha_n})\]
where, player $i$ chooses to play the strategy $s_i^{\alpha_i}$ ($\alpha_i \in \{1, 2, \ldots, \abs{\mathcal{S}_i}\}$). Then, depending on this play, each player $j$ receives a reward that is computed by the pay-off function $\pi_j$ - 
\[\pi_j(\mathcal{P}) = \pi_j(s_1^{\alpha_1}, s_2^{\alpha_2}, \ldots, s_n^{\alpha_n})\]
In general, the pay-off that player $j$ receives, depends not only on his/her own strategy $s_j^{\alpha_j}$ but also on the strategies that the other players have played. The analysis of games played by rational players is carried out under the assumption that, each player tries to maximize his/her own payoff. 

In games of ``\emph{Complete Information}'', each player knows the strategy spaces $\mathcal{S}_i$ and payoff functions $\pi_i$ of all the other players. However, a player $i$ doesn't know what strategy player $j$ will use until the game is played and all the players have revealed their strategies. Every player only knows the set of possible strategies that each player can draw from. 

Within this framework, it is interesting to analyze what strategies the players might use. These play of strategies that rational players might use in a game are called `solutions' of the game. One such solution concept is that of a \emph{Dominant Strategy}. 
Player $i$ has a dominant strategy $s_i^D$, if 
\[\pi_i(s_1^{\alpha_1}, \ldots, s_{i-1}^{\alpha_{i-1}}, s_i^D, s_{i+1}^{\alpha_{i+1}}, \ldots s_n^{\alpha_n}) \geq \pi_i(s_1^{\alpha_1}, \ldots, s_{i-1}^{\alpha_{i-1}}, s_i^{\alpha_i}, s_{i+1}^{\alpha_{i+1}}, \ldots s_n^{\alpha_n})\]
\[\forall s_j^{\alpha_j}\in \mathcal{S}_j,\; j = 1, 2, \ldots, i, \ldots, n\qquad \text{for a fixed }i\]
Simply put, a dominant strategy $s_i^D$ maximizes the payoff of player $i$ for \emph{any} play of strategies from the other players. The strategy $s_i^D$ is called \emph{Strictly Dominant} if the above inequality is strict. 

In general, there exist games where some or none of the players have a dominant strategy. However, if any player does have such a strategy, rationality dictates that the player will play the dominant strategy since it maximizes his payoff no matter what the other players do. If \emph{each} player $j$ has a dominant strategy $s_j^{D_j}$, then the play of dominant strategies $\mathcal{P}^D$ - 
\[\mathcal{P}^D = (s_1^{D_1}, s_2^{D_2}, \ldots, s_n^{D_n})\]
can describe a ``solution'' to the game, as in rational players will tend to play this strategy. 

However, as stated earlier, in a general scenario some or none of the players may have a dominant strategy. What can be the solution to the game in that scenario ? We need a more general solution concept to describe these settings. One such solution concept is the famous \emph{Nash Equilibrium} \cite{nash1950equilibrium}. 

A play of strategies $\mathcal{P}^{NE} = (s_1^{e_1}, s_2^{e_2}, \ldots, s_n^{e_n})$ is called a Nash Equilibrium if and only if - 
\[\pi_i(s_1^{e_1}, \ldots, s_{i-1}^{e_{i-1}}, s_i^{e_i}, s_{i+1}^{e_{i+1}}, \ldots s_n^{e_n}) \geq \pi_i(s_1^{e_1}, \ldots, s_{i-1}^{e_{i-1}}, s_i^{\alpha_i}, s_{i+1}^{e_{i+1}}, \ldots s_n^{e_n})\]
\[\forall s_i^{\alpha_i}\in \mathcal{S}_i \; \text{for a fixed }i\]
And this holds for all players $i = 1, 2, \ldots, n$. This means that, \emph{unilateral} deviance of any \emph{single} player from $\mathcal{P}^{NE}$ cannot increase his payoff. So, none of the players have an incentive to deviate from the equilibrium solution. If the above inequality is strict, then the solution is called a \emph{Strict Nash Equilibrium}. A game might have multiple Nash Equilibria. 

As can be seen clearly, if $\mathcal{P}^D$ is a dominant solution to a game \textbf{G}, then it is also a Nash Equilibrium. However, Pure-Strategic games do not always admit a Nash Equilibrium solution. In order to formulate a solution concept that always exists, we have to extend our notion of pure strategies to ``mixed'' strategies. 
 A \emph{Mixed Strategy} game $\textbf{G}_M(\vb{P}_{\mathcal{S}}, \Bar{\pi})$ corresponding to a pure-strategy game \textbf{G}$(\mathcal{S}, \pi)$ is described by a tuple of probability distribution spaces - 
 \[\vb{P}_{\mathcal{S}} = (\vb{P}_{\mathcal{S}_1}, \vb{P}_{\mathcal{S}_2}, \ldots, \vb{P}_{\mathcal{S}_n})\]
 where, 
 \[\vb{P}_{\mathcal{S}_i} = \left\{\vb{p}_i = [p_i^{\alpha_1}\; p_i^{\alpha_2}\; \ldots \; p_i^{\alpha_{\abs{\mathcal{S}_i}}}]\,\Big|\, p_i^{\alpha_i} \geq 0, \sum_{\alpha}p_i^{\alpha_i} = 1\right\}\]
 is the set of all probability distributions over the strategy space $\mathcal{S}_i$. A play of $\vb{G}_M$ now consists of a tuple of probability vectors $\vb{p}_i$ - 
 \[\mathcal{P}_{\vb{G}_M} := (\vb{p}_1, \vb{p}_2, \ldots, \vb{p}_n)\]
 where each $\vb{p}_i$ is a probability distribution over the strategies $s_i^{\alpha_i}$.\\
 The payoff functions $\pi_i$ are now replaced by \emph{expected} payoff functions $\Bar{\pi}_i$, where, 
 \[\Bar{\pi}_i = \sum_{\alpha_1,  \alpha_2, \ldots, \alpha_n} (p_1^{\alpha_1}\cdot p_2^{\alpha_2}\cdot \ldots \cdot p_n^{\alpha_n})\cdot \pi_i(s_1^{\alpha_1}, s_2^{\alpha_2}, \ldots, s_n^{\alpha_n})\]
 which can be succinctly written as - 
 \[\Bar{\pi}_i = \sum_{\mathcal{P}_{\vb{G}}}p(\mathcal{P_{\vb{G}}})\cdot\pi_i(\mathcal{P_{\vb{G}}})\]
 where the sum is over all plays $\mathcal{P}_{\vb{G}}(\vb{\alpha})$ of the \emph{pure-strategy} game $\vb{G}$, and $$p(\mathcal{P}_{\vb{G}}(\vb{\alpha})) = p_1^{\alpha_1}\cdot p_2^{\alpha_2}\cdot \ldots \cdot p_n^{\alpha_n}$$ is the probability that the play $\mathcal{P}_{\vb{G}}(\mathbf{\alpha})$ will occur in the Mixed-Strategy game $\vb{G}_M$.\\
 As described above, the probability distribution $p(\mathcal{P}_{\vb{G}})$ is separable, and each player chooses his/her strategy (viz. probabilities) independently of the other players. We can easily envision probability distributions for $p(\mathcal{P}_{\vb{G}})$ which are not separable, and contemplate whether this results in additional properties of the game or additional equilibria. Such games where $p(\mathcal{P})$ is not separable are called (classically) ``Correlated Games''. The emphasis on the \emph{classical} correlation is because, in the formulation of Quantum Games, we will see that Quantum Mechanics, particularly Quantum Entanglement, can allow for much more exotic correlations which cannot be mimicked by classical means\footnote{Interestingly, this topic was the subject of the 2022 Nobel Prize in Physics.}. 
 
 A remarkable result in the field of Game Theory, proved by Nash \cite{nash1950equilibrium}, is that in any finite mixed-strategy game, there exists at least one Nash Equilibrium. Thus, we have finally arrived at our desired solution concept. 
 
 Before moving on to the formulation of Quantum Games, we review another game-theoretic concept called \emph{Pareto-Dominance} that is relevant in describing the advantage of Quantum Strategies. A play $\mathcal{P}$ of a game $\vb{G}$ (mixed or pure) is \emph{Pareto-Comparable} to another play $\mathcal{P}'$ if -
 \[\pi_i(\mathcal{P}) \geq \pi_i(\mathcal{P}')\; \forall i \qquad \text{or}\qquad \pi_i(\mathcal{P}) \leq \pi_i(\mathcal{P}') \;\forall i\]
 In general, two plays $\mathcal{P}$ and $\mathcal{P'}$ are \emph{not} Pareto-Comparable. However, if they \emph{are} Pareto-Comparable, then, the play $\mathcal{P}$ is said to \emph{Pareto-Dominate} the play $\mathcal{P}'$ if 
 \[\pi_i(\mathcal{P}) \geq \pi_i(\mathcal{P}') \; \forall i\]
 We write this succinctly as - 
 \[\vb{\pi}(\mathcal{P})\succeq \pi(\mathcal{P}')\]
 If the inequality is strict, then $\mathcal{P}$ \emph{strictly} Pareto-dominates $\mathcal{P}'$, and we write it as 
 \[\pi(\mathcal{P})\succ \pi(\mathcal{P}')\]
 A play $\mathcal{P}_o$ is called \emph{Pareto-Optimal} if, for any other play $\mathcal{P}$ comparable to $\mathcal{P}_o$, 
 \[\pi(\mathcal{P}_o) \succeq \pi(\mathcal{P})\]
 This implies that, if the play $\mathcal{P}_o$ is Pareto-Optimal, no other play of strategies can increase the payoff of one player without decreasing the payoff of another. \\Pareto-optimality captures the concept of ``social welfare'', in a way that, it ranks plays in an order which maximizes the overall payoff of all the players. As a neutral judge of a game, we would like to have a solution of a game which is a Nash Equilibrium and which is also Pareto-Optimal. However, a certain solution might not always possess these two criteria at the same time.

\section{Quantum vs. Classical Players}
We begin our discussion of Quantum Game Theory by reviewing one of the seminal results of this field, which initiated the study of Quantum Games. In 1999, David Meyer first proposed that one can construct 2-player games with quantum systems in which, if one player is restricted to Classical Strategies, while the other player has access to Quantum Strategies, the Quantum player will always win \cite{meyer1999quantum}. What is meant by these Classical and Quantum strategies, and the game that Meyer used to demonstrate this result is described below. 

\paragraph{The (Classical) Penny-Flip Game} This is a 2-player game consisting of players $Q$ and $C$, and a coin. The coin starts out in the state Heads. First, Player Q is asked to select between one of 2 moves or strategies - Flip (F) or No-Flip (N), i.e, Q can choose to either Flip the coin or leave it as it is. Next, Player C is asked to do the same, i.e, choose between F or N, without revealing the coin to him/her. Finally, Player Q is again asked to choose between one of the 2 moves. The coin is revealed to both players at the end of these 3 moves. (The coin remains hidden until the 3 moves are over. Each player can only perform operations on the coin, but cannot know its state during the game.) If at the end of the 3 moves, the state of the coin is Heads, Q wins. If it is Tails, C wins. 

The payoff function $\pi_C$ of player $C$ in the Penny-Flip game can be presented in normal form as in Table \ref{table_PQ}.
\begin{table}[!h]
\centering
\begin{tabular}{cccccc}
                                        &                        & \multicolumn{4}{c}{Q}                                                                                 \\ \cline{2-6} 
\multicolumn{1}{c|}{}                   & \multicolumn{1}{l|}{}  & \multicolumn{1}{l|}{NN} & \multicolumn{1}{l|}{NF} & \multicolumn{1}{l|}{FN} & \multicolumn{1}{l|}{FF} \\ \cline{2-6} 
\multicolumn{1}{c|}{\multirow{2}{*}{C}} & \multicolumn{1}{l|}{N} & \multicolumn{1}{l|}{-1} & \multicolumn{1}{l|}{1}  & \multicolumn{1}{l|}{1}  & \multicolumn{1}{l|}{-1} \\ \cline{2-6} 
\multicolumn{1}{c|}{}                   & \multicolumn{1}{l|}{F} & \multicolumn{1}{l|}{1}  & \multicolumn{1}{l|}{-1} & \multicolumn{1}{l|}{-1} & \multicolumn{1}{l|}{1}  \\ \cline{2-6} 
\end{tabular}
\label{table_PQ}
\caption{Payoffs of Player C for different plays of the Penny-Flip Game.}
\end{table}

For any play $\mathcal{P}$, the pay-off of player Q is $\pi_Q(\mathcal{P}) = -\pi_C(\mathcal{P})$. So, as it is currently formulated, this is a \emph{Zero-Sum} game. 

As can be easily checked for this game, none of the players have a dominant strategy and there are no Nash-Equilibria in Pure Strategies. However, there is a Nash Equilibrium in mixed strategies $\mathcal{P}^{NE}(\vb{p}_C, \vb{p}_Q)$, where,
\[\vb{p}_C = \left[\frac{1}{2} \; \frac{1}{2}\right] \qquad \vb{p}_Q = \left[\frac{1}{4} \; \frac{1}{4} \; \frac{1}{4} \; \frac{1}{4}\right]\]
i.e, each player plays each of his/her pure strategies with equal probability with an expected payoff of $\Bar{\pi}_P = \Bar{\pi}_Q = 0$. None of the players can do any better by unilaterally diverting from this strategy. 

\paragraph{The Quantum Penny-Flip Game} Now we move on to ``\emph{Quantumize}'' this game. To do this, we will consider the scenario where the penny in the \emph{Quantum} Penny-Flip game is not a classical penny, but a quantum penny. That is, where a classical penny has two states \textbf{H} (Heads) and \textbf{T} (Tails), a quantum penny is a two-dimensional quantum system with orthonormal basis states $\ket{\textbf{H}} = \ket{0}$ and $\ket{\textbf{T}}=\ket{1}$. The Quantum Penny allows for a much larger state space, since its state $\ket{\psi}$ can be any superposition of the two basis states - 
\[\ket{\psi} = \alpha\ket{\textbf{H}} + \beta\ket{\textbf{T}} \]
The basis states $\ket{\vb{H}}$ and $\ket{\vb{T}}$ correspond to the classical states $\vb{H}$ and $\vb{T}$. The classical strategies of Flip and No-Flip now correspond to Quantum Operators \textsf{F} and \textsf{N}, such that, 
\begin{equation}\label{eq_meyer_class}
    \textsf{F} = \begin{bmatrix}
    0 & 1\\
    1 & 0
\end{bmatrix} \qquad \textsf{N} = \begin{bmatrix}
    1 & 0\\
    0 & 1
\end{bmatrix}
\end{equation}

One can easily check that, $\textsf{F}$ and $\textsf{N} $ are Unitary and satisfy, 
\begin{align*}
    \textsf{F}\ket{\vb{H}} = \ket{\vb{T}} & \, & \textsf{N}\ket{\vb{H}} = \ket{\vb{H}} \\ 
    \textsf{F}\ket{\vb{T}} = \ket{\vb{H}} & \, & \textsf{N}\ket{\vb{T}} = \ket{\vb{T}}
\end{align*}
So, \textsf{F} and \textsf{N} indeed correspond to the classical strategies of Flip and No-Flip. Our quantumization of the Penny-Flip game is complete. 

Now, to demonstrate the advantage of Quantum Strategies, we will allow the player Q (henceforth called the Quantum Player) to have access to \emph{any} unitary operation $U$ to apply to the quantum penny (in addition to $\textsf{F}$ and $\textsf{N}$) during his move, while the Player C (henceforth called the Classical Player) is only restricted to the ``classical'' moves $\textsf{F}$ and $\textsf{N}$. Thus, Q has access to the entire suite of Quantum Operations, a.k.a strategies, on the quantum penny, whereas, C has access to only classical strategies. In this framework, we will now show that, the Quantum Player always wins, i.e, Q can select operators such that, at the end of the 3 moves, the state of the penny is always $\ket{\vb{H}}$ no matter what move C chooses to play. 

The strategy that Q uses is as follows. The quantum penny starts out in the state $\ket{\vb{H}}$. In his first move, Q acts on the penny with unitary operator $\textsf{U}_Q^{\star}$ defined as -  
\[\textsf{U}_Q^{\star} = \frac{1}{\sqrt{2}}\begin{bmatrix}
    1 & 1\\
    1 & -1
\end{bmatrix}\]
After this, the penny is passed to player C who applies an operators $\textsf{U}_C \in \{\textsf{F}, \textsf{N}\}$ to it. Next the penny is passed back to player Q, who again applies the operator $\textsf{U}_Q^{\star}$ to it, as defined above. So, the final state of the penny is $\textsf{U}_Q^{\star}\textsf{U}_C\textsf{U}_Q^{\star}\ket{\vb{H}}$. The judge of the game now performs a measurement on the penny using the projection operators $\{\dyad{\vb{H}}, \dyad{\vb{T}}\}$. If the measurement outcome is $\ket{\vb{H}}$, Q wins. Otherwise, if it is $\ket{\vb{T}}$, C wins. 

The remarkable thing is that, one can easily calculate, 
\[\textsf{U}_Q^{\star}\textsf{F}_C\textsf{U}_Q^{\star}\ket{\vb{H}} = \ket{\vb{H}}\]
\[\textsf{U}_Q^{\star}\textsf{N}_C\textsf{U}_Q^{\star}\ket{\vb{H}} = \ket{\vb{H}}\]
So, no matter what strategy the classical player uses (given that he is restricted to the classical strategies $\textsf{F}$ and $\textsf{N}$), if the Quantum Player uses the operator $\textsf{U}_Q^{\star}$ defined above, the final state of the penny is always $\ket{\vb{H}}$, and thus, Q always wins. 

This happens because, 
\[\textsf{U}_Q^{\star}\ket{\vb{H}} = \frac{1}{\sqrt{2}}\ket{\vb{H}} + \frac{1}{\sqrt{2}}\ket{\vb{T}} = \ket{+}\]
One can check that, the $\ket{+}$ state is an eigenvector of both the operators \textsf{F} and \textsf{N}, with an eigenvalue of 1, i.e, 
\[\textsf{F}\ket{+} = \ket{+}\qquad \textsf{N}\ket{+} = \ket{+}\]
So, no matter which operation player C chooses in the 2nd move, the $\ket{+}$ state remains unchanged. Then, in the 3rd move, player Q can again use the operation $\textsf{U}_Q^{\star}$ to revert the $\ket{+}$ state back to the $\ket{\vb{H}}$ state (because $\textsf{U}_Q^{\star}\ket{+} = \ket{\vb{H}}$), thus always ensuring a win. This is the underlying mechanism why Quantum Strategies give an edge to the player Q in this game.

\section{Quantum vs. Quantum Players}
As we saw in the previous section, in a Quantum Game (one in which quantum systems are manipulated and communication is done utilizing quantum technologies), if a player restricts himself to only Classical Strategies, he risks to always losing to a player utilizing Quantum Strategies. So, rationality dictates that, all the players in a quantum game will try to use quantum strategies to have a level playing field. 

Next, we quantumize another game in way that both players have access to quantum strategies. We will see if they can utilize their strategies to obtain a better payoff than in the classical case. 

\paragraph{The Prisoner's Dilemma} This is perhaps one of the most famous and extensively studied game in Game Theory. There are 2 prisoners A and B, who are suspected of committing a crime. Each prisoner is asked by a judge if the other has committed the crime. If one prisoner rats the other one out (Defects), while the other one remains silent (Cooperates), the cooperative player will receive a sentence of $\alpha$ years and the defective player will be set free. However, if both players defect, they will each receive a sentence of $\beta$ years ($\beta < \alpha$). And, if both cooperate, they will each receive a sentence of $\gamma$ years ($\gamma < \beta$). 

We can dexcribe this scenario by the following payoff bi-matrix - 

\begin{table}[!h]
\centering
\begin{tabular}{lccc}
                                        & \multicolumn{1}{l}{}     & \multicolumn{2}{c}{B}                                                               \\ \cline{2-4} 
\multicolumn{1}{l|}{}                   & \multicolumn{1}{c|}{}    & \multicolumn{1}{c|}{$C$}                  & \multicolumn{1}{c|}{$D$}                \\ \cline{2-4} 
\multicolumn{1}{c|}{\multirow{2}{*}{A}} & \multicolumn{1}{c|}{$C$} & \multicolumn{1}{c|}{$(-\gamma, -\gamma)$} & \multicolumn{1}{c|}{$(-\alpha, 0)$}     \\ \cline{2-4} 
\multicolumn{1}{c|}{}                   & \multicolumn{1}{c|}{$D$} & \multicolumn{1}{c|}{$(0, -\alpha)$}       & \multicolumn{1}{c|}{$(-\beta, -\beta)$} \\ \cline{2-4} 
\end{tabular}
\label{table_pris_dil}
\end{table}

where, the first element of a tuple represents A's payoff, and the second element represents B's.

In this game, we can see that Defection (D) is a dominant strategy for both players. I.e, no matter what the other player does, defection results in a greater payoff for a particular player. This results in a dominant solution $\mathcal{P}^D = (D, D)$ for the game. In addition, $(D, D)$ is also a Nash Equilibrium, i.e., if one player is adamant at defecting, the other player is worse-off if he chooses to cooperate. However, we can clearly see that, $(D, D)$ is  \emph{suboptimal} to the Pareto-Optimal play $(C, C)$, where each player can get a higher payoff if they both choose to cooperate rather than defect. Except the dominant play, there are no other (pure or mixed) Nash Equilibrium. 

\paragraph{The Quantum Prisoner's Dilemma} Now, we move on to quantumize the Prisonner's Dillema game. We follow the scheme proposed by Eisert, Wilkens and Lewenstein (EWL) \etal \cite{eisert1999quantum, eisert2000quantum} which will allow us to generalize and quantumize arbitrary Classical Games.

% corresponding to the classical strategies of C (Cooperation) and D (Defection). Note that, in Meyer's approach \cite{meyer1999quantum}, the classical strategies were mapped to operators on the Hilbert Space of a Quantum System. Whereas, in the EWL approach, they are mapped to quantum states themselves.

First, we choose 2 two-dimensional Quantum Systems A and B, with basis states $\ket{{0}}$ and $\ket{{1}}$. A judge prepares the composite system AB in a state $\ket{\psi}$, which can be entangled in general. Suppose, it starts out in the state - 
\[\ket{\eta_{in}}_{AB} = \frac{1}{\sqrt{2}}\ket{00} + i\frac{1}{\sqrt{2}}\ket{11}\]
($\ket{00}$ stands for $\ket{0}_A\otimes \ket{0}_B$. We omit the explicit tensor products $\otimes$ between the states for simplicity, and the labels A and B when they are evident from context.) Next, the two systems $A$ and $B$ are sent to the players A and B. Each player then applies operations $U_A$ and $U_B$ on their sub-systems and returns them to the judge. The composite state that is returned to the judge is - 
\[\ket{\eta_{f}}_{AB} = (U_A\otimes U_B)\ket{\eta_{in}}_{AB}\]
Then, the judge performs a projective measurement on the entire system $AB$ in the basis $\{\ket{\eta_{CC}}, \ket{\eta_{CD}}, \ket{\eta_{DC}}, \ket{\eta_{DD}}\}$, where, 
\begin{align*}
    \ket{\eta_{CC}} &= (\ket{00} + i\ket{11})/\sqrt{2} & \ket{\eta_{CD}} &= (\ket{01} - i\ket{10})/\sqrt{2}\\
     \ket{\eta_{DD}} &= (\ket{00} - i\ket{11})/\sqrt{2} & \ket{\eta_{DC}} &= (\ket{01} + i\ket{10})/\sqrt{2}
\end{align*}

using the orthonormal projectors $\Pi_{CC} = \dyad{\eta_{CC}}$, etc. Each player then receives a payoff depending on the outcome of the measurement, where a measurement outcome of $\ket{\eta_{CC}}$ corresponds to a play of the Classical Strategy $(C, C)$, and thus, each of the projectors correspond to particular plays of the pure strategy game.  

An important point to note here is that, in Meyer's scheme \cite{meyer1999quantum}, the Classical Strategies are mapped to Unitary Operators on a suitable Hilbert Space ($\vb{F}\rightarrow \textsf{F}, \vb{N}\rightarrow \textsf{N}$). On the other hand, in the EWL scheme, they are mapped to orthonormal projectors of a projective measurement ($(C, C) \rightarrow \Pi_{CC}$, etc.) These are two different quantumization schemes, and we will see that the EWL scheme is more general and can be used to quantumize any Classical Game into a Quantum Correlated game, whereas Meyer's scheme is applicable only to a certain special class of games where the players manipulate a common object. 

Now, we continue with our discussion of the EWL quantumization scheme. Since the measurement outcome of the Quantum Game is probabilistic, the expected payoff of a player, suppose A, is calculated as - 
\begin{align*}
    \Bar{\pi}_A &= p(CC)\pi_A(CC) + p(CD)\pi_A(CD) + p(DC)\pi_A(DC) + p(DD)\pi_A(DD)\\
    &= \bra{\eta_{f}}\Pi_{CC}\ket{\eta_{f}}\cdot\pi_A(CC) + \bra{\eta_{f}}\Pi_{CD}\ket{\eta_{f}}\cdot\pi_A(CD)\\
    & \quad + \bra{\eta_{f}}\Pi_{DC}\ket{\eta_{f}}\cdot\pi_A(DC)  + \bra{\eta_{f}}\Pi_{DD}\ket{\eta_{f}}\cdot\pi_A(DD)\\
    & = \bra{\eta_{f}}\Big(\underbrace{\Pi_{CC}\pi_A(CC) + \Pi_{CC}\pi_A(CC) + \Pi_{CC}\pi_A(CC) + \Pi_{CC}\pi_A(CC)}_{\hat{\pi}_A}\Big) \ket{\eta_{f}}\\
    &= \bra{\eta_{f}}\hat{\pi}_A\ket{\eta_{f}}
\end{align*}

where, we define $\hat{\pi}_A$ above as player A's \emph{Payoff Operator}. 

Similarly, $\Bar{\pi}_B = \bra{\eta_{f}}\hat{\pi}_B\ket{\eta_{f}}$, with $\hat{\pi}_B$ defined similarly as above. 

A play of the Quantum Prisoner's Dilemma thus consists of a tuple $(U_A, U_B)$ of the operations that A and B choose to use on an initial state $\ket{\eta_{in}}$, with the expected payoffs defined as above. This is similar to the framework of Correlated Games, where individual players draw their strategies from a joint probability distribution. In the quantum version, the operations $U_A$ and $U_B$ locally modify the probability distribution that finally results from the judge's measurement. However, initial quantum entanglement allows for much more exotic correlations than is possible to achieve from a classical joint probability distribution.

Now, we analyze what are the equilibria of this game under different restrictions for $U_A, U_B$. 

\subsubsection*{One Parameter Strategies}

First, we consider $U$'s of the form - 
\[U(\theta) = \begin{bmatrix}
    \cos{(\theta/2)} & \sin{(\theta/2)}\\
    -\sin{(\theta/2)} & \cos{(\theta/2)}
\end{bmatrix}\]
where, $\theta \in [0, \pi]$.
Here, if we define 
\[U_C = U(0) \qquad \text{and} \qquad U_D = U(\pi)\]
then, we can easily calculate that, 
\[(U_{\alpha}\otimes U_{\beta})\ket{\eta_{in}} = \ket{\eta_{\alpha\beta}}\qquad \alpha, \beta \in \{C, D\}\]
In general, 
\begin{align*}
    \ket{\eta_f} &=(U(\theta_A)\otimes U(\theta_B)\ket{\eta_{in}} \\
    &= \cos{(\theta_A/2)} \cos{(\theta_B/2)} \ket{\eta_{CC}} + i \sin{(\theta_A/2)} \sin{(\theta_B/2)} \ket{\eta_{DD}}\\
    & \;\;- \cos{(\theta_A/2)} \sin{(\theta_B/2)} \ket{\eta_{CD}} + i \sin{(\theta_A/2)} \cos{(\theta_B/2)} \ket{\eta_{DC}} 
\end{align*}
where, we have used the following formula to arrive at the above equality - 
\[U(\theta)\ket{0} = \cos{(\theta/2)}\ket{0} - \sin{(\theta/2)}\ket{1} \qquad U(\theta)\ket{1} = \sin{(\theta/2)}\ket{0} + \cos{(\theta/2)}\ket{1}\]
And, the change of basis formula - 
\begin{align*}
    \ket{00} &= (\ket{\eta_{CC}} + \ket{\eta_{DD}})/\sqrt{2} & \ket{01} &= (\ket{\eta_{DC}} + \ket{\eta_{CD}})/\sqrt{2}\\
     \ket{11} &= (\ket{\eta_{CC}} - \ket{\eta_{DD}})/\sqrt{2}i & \ket{10} &= (\ket{\eta_{DC}} - \ket{\eta_{CD}})/\sqrt{2}i
\end{align*}
From the final form of $\ket{\eta_f}$, the expected payoffs are - 
\[\Bar{\pi}_A = -\gamma\cdot\abs{\cos{(\theta_A/2)} \cos{(\theta_B/2)}}^2 - \beta\cdot \abs{\sin{(\theta_A/2)} \sin{(\theta_B/2)}}^2 - \alpha \abs{\cos{(\theta_A/2)} \sin{(\theta_B/2)}}^2\]
and, similarly for $\Bar{\pi}_B$. 

The payoff functions are exactly equivalent to a classical mixed strategy Prisoner's Dilemma game with probability $p=\cos^2(\theta/2)$ to Cooperate. The players $A$ and $B$ each try to convert the state $\ket{\eta_{in}}$ to the state $\eta_{DC}$ and $\eta_{CD}$ respectively, to obtain maximum payoff. 
So, just like the classical game, each player has a dominant strategy which is to defect (i.e, play $U_D$). However, once a player plays $U_D$, the best option for the other player is to play $U_D$ as well, which  modifies the initial state to $\ket{\eta_{DD}}$. This results in a measurement outcome of $(D, D)$, and thus the players settle in the dominant but Pareto-suboptimal Nash Equilibrium. This case doesn't provide any advantage over the classical case. 

\subsubsection*{Two Parameter Strategies}
Next, we extend the set of allowed strategies to the 2-parameter Unitary Operators described below: 
\[U(\theta, \phi) = \begin{bmatrix}
    e^{i\phi}\cos{(\theta/2)} & \sin{(\theta/2)}\\
    -\sin{(\theta/2)} & e^{-i\phi}\cos{(\theta/2)}
\end{bmatrix}\]
with $\theta \in [0, \pi]$ and $\phi \in [0, \pi/2]$. 

As before, the classical pure strategies correspond to the operators - 
\[U_C = U(0, 0) \qquad \text{and} \qquad U_D = U(\pi, 0)\]
As it stands, the strategy $U_D$ is no longer a dominant strategy. If one player plays $U_D$, the other player can maximize his/her payoff by deviating from $U_D$. For example, if player B plays $U_D$, we have,
\[(I\otimes U_D)\ket{\eta_{in}} = \ket{\eta_{CD}}\]
However, in the space of 2-parameter strategies, $A$ can now use the operation 
\[U_Q = U(0, \pi/2) = \begin{bmatrix}
    i & 0\\
    0 & -i
\end{bmatrix}\]
which results in the state (after B's play of $U_D$),
\[(U_Q\otimes U_D)\ket{\eta_{in}} = (U_Q\otimes I)\ket{\eta_{CD}} = \ket{\eta_{DC}}\]
Since, $\ket{\eta_{DC}}$ results in the maximum payoff for player $A$, $U_Q$ is the best response to $U_D$. 
So, $(D, D)$ is no longer a Nash Equilibrium.

To find a new Nash Equilirbium, we write the general state after $A$ and $B$'s operations $U(\theta_A, \phi_A)$ and $U(\theta_B, \phi_B)$, $\ket{\eta_f} = U(\theta_A, \phi_A) \otimes U(\theta_B, \phi_B) \ket{\eta_{in}}$, which gives after some careful calculation - 
\begin{align*}
     \ket{\eta_f} &= \cos{(\phi_A + \phi_B)}\cos{(\theta_A/2)}\cos{(\theta_B/2)}\ket{\eta_{CC}} \\
    &+ i\left[ \sin{(\theta_A/2)} \sin{(\theta_B/2)} + \sin(\phi_A + \phi_B) \cos{(\theta_A/2)} \cos{(\theta_B/2)} \right]\ket{\eta_{DD}}\\
    &+i\left[\cos{(\phi_B)}\sin{(\theta_A/2)}\cos{(\theta_B/2)} - \sin{(\phi_A)}\cos{(\theta_A/2)}\sin{(\theta_B/2)}\right]\ket{\eta_{DC}}\\
    &+ \left[\sin{(\phi_B)}\sin{(\theta_A/2)}\cos{(\theta_B/2)} - \cos{(\phi_A)}\cos{(\theta_A/2)}\sin{(\theta_B/2)}\right]\ket{\eta_{CD}}
\end{align*}
From the expression of $\ket{\eta_f}$, we can calculate  the expected Payoff of $A$ as - 
\begin{align*}
    \Bar{\pi}_A = &-\alpha\cdot \abs{\sin{(\phi_B)}\sin{(\theta_A/2)}\cos{(\theta_B/2)} - \cos{(\phi_A)} \cos{(\theta_A/2)} \sin{(\theta_B/2)}}^2\\
    &- \beta\cdot \abs{\cos{(\phi_A + \phi_B)}\cos{(\theta_A/2)}\cos{(\theta_B/2)}}^2\\
    &- \gamma\cdot \abs{\sin{(\theta_A/2)} \sin{(\theta_B/2)} + \sin(\phi_A + \phi_B) \cos{(\theta_A/2)} \cos{(\theta_B/2)}}^2
\end{align*}
And, similarly for $\Bar{\pi}_B$. For any play $(\theta_B, \phi_B)$ of $B$, we can optimize this expression by setting - 
\[\pdv{\Bar{\pi}_A}{\theta_A} = 0\qquad \text{and}\qquad\pdv{\Bar{\pi}_A}{\phi_A} = 0\]
From this, we get, 
\begin{align}
    \tan(\theta_A/2)\cos{(\phi_B)} &+ \tan(\theta_B/2)\cos{(\phi_A)} = 0 \\
    \tan(\theta_A/2)\sin{(\phi_B)} &+ \cot(\theta_B/2)\cos{(\phi_A)} = 0
\end{align}
which can be solved for a fixed $(\theta_B, \phi_B)$ to find the optimal $(\theta_A, \phi_A)$. For example, if $(\theta_B, \phi_B) = (\pi, 0)$, the optimal move for $A$ is $(\theta_A, \phi_A) = (0, \pi/2)$ which results in the operator $U_Q$ defined above. 

Particularly, we can calculate the best response for the move $(\theta_B, \phi_B) = (0, \pi/2)$ (i.e, $U_Q$) to be - 
\[\mathcal{BR}_A(0, \pi/2) = \{(\theta_A, \phi_A)\; \big|\; \theta_A\in [0, \pi];\; \phi_A = \pi/2\}\]
So, $(0, \pi/2) \in \mathcal{BR}_A(0, \pi/2)$, and so, $(U_Q, U_Q)$ is a Quantum Nash Equilibrium for the Quantum Prisoner's Dilemma in the set of 2-parameter of strategies. 

We can easily calculate that, $\Bar{\pi}_A(U_Q, U_Q) = \Bar{\pi}_B(U_Q, U_Q) = -\gamma$, which is the Pareto-Optimal solution. So, in this case, Quantum Strategies allow players to realize the Pareto-Optimal play as a Nash Equilibrium, which is not possible in the classical game.\\

So far, we have considered Pure Strategy Quantum Games where the players only apply \emph{one} of the Unitary Operators they have access to. We can extend this notion further to \emph{Mixed Strategy Quantum Games}, where the initial state of the Quantum System can be mixed, and the players can apply their strategies probabilistically. To do that, we have to extend our description of Quantum Mechanics to account for this stochasticity.

\subsubsection*{Three Parameter Strategies}
Now, let's analyze the most general case, where $U_A, U_B$ can be any allowable Unitary Operator. In this case, the operators can be described by 3 parameters as follows - 
\[U(\theta, \phi, \lambda) = \begin{bmatrix}
    e^{i\phi}\cos{(\theta/2)} & e^{-i\lambda}\sin{(\theta/2)}\\
    -e^{i\lambda}\sin{(\theta/2)} & e^{-i\phi}\cos{(\theta/2)}
\end{bmatrix}\]
As it follows, there exists no Nash Equilibria in this set of strategies. Because, for any strategy $U_A = U(\theta_A, \phi_A, \lambda_A)$, B can choose a strategy $U_B$ such that, 
\[(U_A\otimes U_B)\ket{\eta_i} = \ket{\eta_{CD}}\]
i.e, B can manipulate the state so that it gives him the maximum payoff at the expense of A. Particularly, B can choose the operator - 
\[U_B = U_DU_A^{\dagger} = U(\pi, 0, 0)\cdot U(-\theta_A, -\phi_A, -\lambda_A)\]
Since, by symmetry, A can do the same thing, there exists no Nash Equilibrium for this set of \emph{pure} Quantum Strategies. However, a Nash Equilibrium \emph{can} be found, if we allow for \emph{Mixed} Quantum Strategies. To do that, we have to extend our description of Quantum Mechanics using another mathematical tool.

\section{The Density Matrix Formalism}
To allow for Mixed Quantum Strategies, we have to find a way to describe probabilistic Quantum States and and probabilistic Quantum Operations. In Quantum Mechanics, the Measurement Postulate is the main vehicle which connects the state-vector description of a system to results observed in real world experiments, by providing a means to calculate the probabilities for different outcomes. Now, we need another quantity that will enable us to calculate measurement probabilities from an ensemble of states. This description is due to von Neumann \cite{neumann1927wahrscheinlichkeitstheoretischer} and Landau \cite{landau1927damping}. 

Suppose, we have an ensemble of states $\mathcal{E}:=\{(q_1, \ket{\psi_1}), (q_2, \ket{\psi_2}), \ldots (q_n, \ket{\psi_n})\}$, where the state $\ket{\psi_i}$ occurs with the probability $q_i$. If we are performing a measurement of the observable $O$ on this ensemble, then the probability that the outcome $o$ will occur (which has a corresponding projector $\Pi_o$) is - 
\begin{align*}
    p(o) &= q_1\cdot p(o \,\big| \ket{\psi_1}) + q_2\cdot p(o \,\big| \ket{\psi_2}) + \ldots + q_n\cdot p(o \,\big| \ket{\psi_n}) \\ 
    &= q_1\cdot \bra{\psi_1}\Pi_o\ket{\psi_1} + q_2\cdot \bra{\psi_2}\Pi_o\ket{\psi_2} + \ldots + q_n \cdot\bra{\psi_n}\Pi_o\ket{\psi_n}\\
    &= q_1\cdot\Tr\left[\Pi_o \dyad{\psi_1}\right] + q_2\cdot\Tr\left[\Pi_o \dyad{\psi_2}\right] + \ldots + q_n\cdot\Tr\left[\Pi_o \dyad{\psi_n}\right]\\
    &= \Tr \Big[\Pi_o \bigg( \underbrace{q_1\dyad{\psi_1}+ q_2\dyad{\psi_2} + \ldots + q_n\dyad{\psi_n}}_{\rho}\bigg)\Big]
\end{align*}
where, $\Tr$ is the Trace Operator. 

The quantity in the parentheses is the desired quantity that will allow us to describe the ensemble of states and extract measurement probabilities from it. This quantity $\rho$ is called the \emph{Density Matrix} of the ensemble. So, the density matrix corresponding to the ensemble $\mathcal{E}$ is - 
\[\rho_{\mathcal{E}} = q_1\dyad{\psi_1}+ q_2\dyad{\psi_2} + \ldots + q_n\dyad{\psi_n} = \sum_i q_i\dyad{\psi_i}\]
Such states $\rho_{\mathcal{E}}$ are called \emph{Mixed} states as opposed to \emph{Pure} states which are of the form $\rho = \dyad{\psi}$.

The recipe to get the measurement probability for an outcome $o$ with projector $\Pi_o$ is -
\[p(o) = \Tr[\Pi_o\rho] = \Tr[\rho\Pi_o]\]
where, in the last equality, we have used the cyclic property of Trace. 

Evolution of states can be described in the Density Matrix formalism as follows. Suppose, a Unitary Operator $U$ acts on the ensemble $\mathcal{E}$. This means that, each state of the ensemble has been transformed to - 
\[\ket{\psi_i} \rightarrow U\ket{\psi_i}\]
Hence, the ensemble $\mathcal{E}$ is transformed to - 
\[\mathcal{E}\rightarrow \mathcal{E'} :=\{(q_1, U\ket{\psi_1}), (q_2, U\ket{\psi_2}), \ldots (q_n, U\ket{\psi_n})\}\]
So, the density matrix of the evolved ensemble is -
\[\rho_{\mathcal{E'}} = \sum_i q_iU\dyad{\psi_i}U^{\dagger} = U\left(\sum_i q_i\dyad{\psi_i}\right)U^{\dagger} = U\rho_{\mathcal{E}}U^{\dagger}\]
Simply put, under a Unitary Transformation, a Density Matrix evolves as - 
\[\rho\longrightarrow \rho' = U\rho U^{\dagger}\]
Using this machinery, we can describe even more general scenarios. For example, if a unitary transformation $U_i$ acts on a density matrix $\rho$ with probability $s_i$, then the resulting Density Matrix is -
\[\rho \longrightarrow \rho' = \sum_i s_i\cdot U_i\rho U_i^{\dagger} = \sum_i E_i\rho E_i^{\dagger}\]
where, 
\[E_i = \sqrt{s_i}U_i\]
are called the `Kraus Operators' of the generalized evolution. 

We can use the language of Desnity Matrices to formulate Mixed Strategy Quantum Games. However, before doing that, we generalize the approaches of Meyer \cite{meyer1999quantum} and EWL \cite{eisert2000quantum} to formulate a protocol to quantumize any Classical Game in this language. We will then use this protocol to formulate a Mixed Strategy Quantum Game due to Marinatto and Weber \cite{marinatto2000quantum}.

\section{Game Quantumization Protocol}
Following the description of EWL, we now describe a general protocol to quantumize any $n$-player finite classical pure-strategy game $\vb{G}(\mathcal{S}, \pi)$ (as desribed in eq. \ref{pG_S}-\ref{pG_strat}) to the pure strategy quantum game $\vb{G}^Q$. The steps go as follows - 
\begin{enumerate}
    \item First, we choose $n$ quantum systems $Q_i$ with dimensions $\abs{\mathcal{S}_i}$ respectively.
    \item A judge prepares the composite system $\mathcal{Q}$, 
    \[\mathcal{Q} = Q_1\otimes Q_2\otimes\ldots\otimes Q_n\]
    in an initial state $\rho$, which might in general be entangled and mixed.  
    \item The judge then passes each of the subsystems $Q_i$ to the player $i$. 
    \item Each player $i$ then operates on $Q_i$ with a Unitary operator $U_i$, and returns their respective subsystem to the judge. The combined unitary operation on the entire system is $\mathcal{U} = U_1\otimes U_2\otimes \ldots \otimes U_n$. So, the state returned to the judge is - 
    \[\rho_f = \mathcal{U} \rho_{in} \mathcal{U}^{\dagger} \]
    The set of operators $U_i$ now represent the strategy-space of the player $i$ in the Quantum Game $\vb{G}^Q$, and the operator $\mathcal{U}$ now represents a play $\mathcal{P}_{\vb{G}^Q}$ of $\vb{G}^Q$. 
    \item The judge then performs a measurement on the state $\rho_f$ of the combined system $\mathcal{Q}$, and gets an outcome corresponding to a play $\mathcal{P}_{\vb{G}}$ of the pure-strategy game $\vb{G}$. The projection operators that the judge uses for measurement can be represented as - 
    \[\Pi_{\vb{G}} = \{\Pi_{\mathcal{P}_{\vb{G}}}\; \big|\; \mathcal{P}_{\vb{G}}:= (s_1^{\alpha_1}, s_2^{\alpha_2}, \ldots, s_n^{\alpha_n}), \forall s_i^{\alpha_i}\in \mathcal{S}_i\}\]
    That is, there is a projector corresponding to each play of pure strategies $\mathcal{P}_{\vb{G}}$. This is consistent with our description of $\mathcal{Q}$, because, the dimension of $\mathcal{Q}$ is, 
    \[\dim(\mathcal{Q}) = \dim(Q_1)\cdot \dim(Q_2) \cdot \ldots \cdot \dim(Q_n) = \abs{\mathcal{S}_1}\cdot \abs{\mathcal{S}_2}\cdot \ldots \cdot \abs{\mathcal{S}_n} \] which is exactly equal to the number of plays of the pure-strategy game $\vb{G}$, and thus, we can always find a complete set of orthonormal projectors equal to this number.
    
    \item Finally, the judge calculates the classical payoffs $\pi_i(\mathcal{P}_{\vb{G}})$ for the play resulting from the measurement, and relays them to the player $i$. Since the outcome of the measurement is probabilistic, the expected payoff of each player is - 
    \[\Bar{\pi}_i(\mathcal{U}) = \sum_{\mathcal{P}_{\vb{G}}} p(\mathcal{P}_{\vb{G}}\, \big| \, \mathcal{U}\rho_{in}\mathcal{U}^{\dagger})\cdot \pi_i(\mathcal{P}_{\vb{G}})\]
\end{enumerate}

Steps 5 and 6 can also be described succinctly using the language of \emph{Payoff Operators}. We can define the payoff operator $\hat{\pi}_i$ for the $i$-th player as - 
\[\hat{\pi}_i = \sum_{\mathcal{P}_{\vb{G}}}\pi_i(\mathcal{P}_{\vb{G}})\cdot\Pi_{\mathcal{P}_{\vb{G}}}\]
As defined, the summation given in the definition is actually the Spectral Decomposition of the operator $\hat{\pi}_i$ (the $\Pi_{\mathcal{P}_{\vb{G}}}$ being orthonormal), and each of the operators $\hat{\pi}_i$ are Hermitian (because the $\pi_i(\mathcal{P}_{\vb{G}})$ are real). As a result, the $\hat{\pi}_i$'s are valid Quantum Observables. In addition, since they share a common eigenbasis, all the operators $\hat{\pi}_i$ commute - 
\[[\hat{\pi}_i, \hat{\pi}_j] = 0\quad \forall\; i, j\]
Using these commutation relations, we can flip the argument for measurement. Given a set of mutually commuting Payoff Operators $\{\hat{\pi}_i\}$, there exists a common eigenbasis for the set of operators. We will measure the state $\mathcal{U}\rho_{in}\mathcal{U}^{\dagger}$ in this common eigenbasis of the $\hat{\pi}_i$'s and calculate the final payoffs from the measurement results. This gives us a way to describe the quantum game, without explicitly mentioning the measurement basis. 

Also, the expected payoff of each player is more concisely expressed using the payoff operators. We simply have, 
\begin{equation}\label{eq_payoff_main}
    \Bar{\pi}_i(\mathcal{U}) = \Tr \left[\mathcal{U}\rho\,\mathcal{U}^{\dagger}\hat{\pi}_i\right] = \Tr \left[\rho\,\mathcal{U}^{\dagger}\hat{\pi}_i\mathcal{U}\right]
\end{equation}

So, the quantum game can be described as $\vb{G}^Q(\rho_{in}, \{\mathcal{U}\}, \{\hat{\pi}_i\})$ with $[\hat{\pi}_i, \hat{\pi}_j] = 0$ for all pairs $i, j$. 
This description can be generalized to \emph{Mixed Quantum Games} $\vb{G}^Q_M(\rho_{in}, \{\mathcal{E}\}, \{\hat{\pi}_i\})$ where the set of Unitary Operators $\{\mathcal{U}\}$ is replaced by the set 
\[\{\mathcal{E}\} = \{\mathcal{E}_1\otimes \mathcal{E}_2\otimes \ldots \otimes \mathcal{E}_n\}\]
where, each $\mathcal{E}_i$ is a generalized Quantum Evolution Operator (which is a Completely Positive Trace Preserving (CPTP) Map on the Reduced Density Matrix $\rho_i = \Tr_{1, 2, \ldots, i-1, i+1, \ldots, n} [\rho]$). In general, $\mathcal{E}_i$ include strategies where the player $i$ selects an operator $\{U_i\}$ with some probability measure $\mu(U_i)$, as well as other more general situations.

\section{A Mixed Strategy Quantum Game}
We end our discussion by using the protocol described in the previous section to quantumize another game. This quantumization was done by Marinatto and Weber \cite{marinatto2000quantum}, but our presentation is much more succinct using the protocol we have developed. 

\paragraph{The Battle of Sexes} In this game, there are two players Alice (A) and Bob (B) who are husband and wife. They want to decide on an activity to do on a certain evening. Alice wants to go outside to watch an Opera (O), while Bob wants to stay at home and watch TV (T). However, they would rather spend time together than alone. This scenario can be described by the following payoff bi-matrix - 
\begin{table}[!h]
\centering
\begin{tabular}{lccc}
                                        & \multicolumn{1}{l}{}   & \multicolumn{2}{c}{B}                                                             \\ \cline{2-4} 
\multicolumn{1}{l|}{}                   & \multicolumn{1}{c|}{}  & \multicolumn{1}{c|}{O}                  & \multicolumn{1}{c|}{T}                  \\ \cline{2-4} 
\multicolumn{1}{c|}{\multirow{2}{*}{A}} & \multicolumn{1}{c|}{O} & \multicolumn{1}{c|}{$(\alpha, \beta)$}  & \multicolumn{1}{c|}{$(\gamma, \gamma)$} \\ \cline{2-4} 
\multicolumn{1}{c|}{}                   & \multicolumn{1}{c|}{T} & \multicolumn{1}{c|}{$(\gamma, \gamma)$} & \multicolumn{1}{c|}{$(\beta, \alpha)$}  \\ \cline{2-4} 
\end{tabular}
\end{table}

where, the first element of a tuple represents Alice's payoff, and the second element represents Bob's. To reflect the preferences of the players in this game, we have $\alpha > \beta > \gamma$. 

This game has 2 pure-strategy Nash Equilibria $\mathcal{P}_1 = (O, O)$ and $\mathcal{P}_2 = (T, T)$ with payoffs $(\alpha, \beta)$ and $(\beta, \alpha)$ respectively. In addition, it also has a mixed strategy Nash Equilibrium $\mathcal{P}_M = (\vb{p}_A, \vb{q}_B)$ with, 
\[\vb{p}_A = \left[\frac{\alpha-\gamma}{\alpha + \beta - \gamma} \quad \frac{\beta-\gamma}{\alpha + \beta - \gamma}\right]\]
\[\vb{q}_B = \left[\frac{\beta-\gamma}{\alpha + \beta - \gamma} \quad \frac{\alpha-\gamma}{\alpha + \beta - \gamma}\right]\]
where the first and second elements of the vectors are the probabilities of selecting $O$ and $T$ for the corresponding players.

This results in an expected payoff of - 
\[\pi_A(\mathcal{P}_M) = \pi_B(\mathcal{P}_M) = \frac{\alpha\beta - \gamma^2}{\alpha + \beta - 2\gamma}\]\\

To quantumize this game, we choose, 
\[\mathcal{Q} = \mathcal{H}_A\otimes \mathcal{H}_B\]
with $\mathcal{H}_A, \mathcal{H}_B$ two-dimensional quantum systems (qubits). 
For demonstration purposes, we choose $\rho_{in} = \dyad{\Phi^+}$ (according to \cite{marinatto2000quantum}), where, 
\[\ket{\Phi^+} = \frac{1}{\sqrt{2}}\ket{00} + \frac{1}{\sqrt{2}}\ket{11}\]
is a maximally entangled state.\\
We restrict the set of operations of $A$ and $B$ to 
\[\{U_A\} = \{U_B\} = \{\textsf{X}, \textsf{I}\}\]
where, 
\[\textsf{X} = \begin{bmatrix}
    0 & 1\\
    1 & 0
\end{bmatrix} \qquad \textsf{I} = \begin{bmatrix}
    1 & 0\\
    0 & 1
\end{bmatrix}\]
Finally, we choose the projective measurement operators 
\[(\Pi_{OO}, \Pi_{OT}, \Pi_{TO}, \Pi_{TT}) = (\dyad{\Phi^+}, \dyad{\Psi^+},\dyad{\Psi^-},  \dyad{\Phi^-})\]
where, 
\begin{align*}
    \ket{\Phi^+} = (\ket{00}+\ket{11})/\sqrt{2} & \quad &\ket{\Psi^+} = (\ket{01}+\ket{10})/\sqrt{2}\\
    \ket{\Phi^-} = (\ket{00}-\ket{11})/\sqrt{2} & \quad & \ket{\Psi^-} = (\ket{01}-\ket{10})/\sqrt{2}
\end{align*}
In this set of strategies, there are 2 pure Quantum Nash Equilibria - 
\[\mathcal{P}^Q_1 = (\textsf{I}, \textsf{I})\qquad \mathcal{P}^Q_2 = (\textsf{X}, \textsf{X})\]
Each with expected payoff - 
\[\Bar{\pi}_A = \Bar{\pi}_B = \frac{\alpha + \beta}{2}\]

Now, we analyze this game for mixed strategies. In this case, player $A$ applies $\textsf{I}$ and $\textsf{X}$ with probabilities $p_{A1} = p_A$ and $p_{A2} = (1-p_A)$, and similarly for $B$, with probabilities $p_{B1} = p_B$ and $p_{B2} = (1-p_B)$. In the mixed strategy game, the final state of $\mathcal{Q}$ is - 
\[\rho_f = \sum_{i, j} p_{Ai}p_{Bj}\cdot (U_{Ai}\otimes U_{Bj})\rho (U_{Ai}^{\dagger}\otimes U_{Bj}^{\dagger})\]
This results in the state, 
\[\rho_f = [p_Ap_B + (1-p_A)(1-p_B)]\dyad{\Phi^+} + [p_A(1-p_B) + (1-p_A)p_B]\dyad{\Psi^+}\]
with expected payoff, 
\begin{align*}
    \Bar{\pi}_A &= \Tr[\rho_f\hat{\pi}_A]\\
    &= [p_Ap_B + (1-p_A)(1-p_B)]\cdot \frac{\alpha + \beta}{2} + [p_A(1-p_B) + (1-p_A)p_B] \cdot \gamma
\end{align*}
So, the Best Response of $B$ for a particular $p_A$ is - 
\[p_B^{\star} = \mathcal{BR}(p_A) = 
\begin{cases}
    0 & p_A > 1/2\\
    1 & p_A < 1/2
\end{cases}\]
However, for $p_A = 1/2$, any $p_B$ will give the same payoff. Particularly, $(p_B=1/2) \in \mathcal{BR}(p_A = 1/2)$. Hence, there exists a Mixed Strategy Quantum Nash Equilibrium, 
\[\mathcal{P}^Q_M = (\left[\frac{1}{2}\;\;\frac{1}{2}\right]\, ,\, \left[\frac{1}{2}\;\;\frac{1}{2}\right])\]
This gives an expected payoff of 
\[\Bar{\pi}_A(\mathcal{P}^Q_M) = \Bar{\pi}_B(\mathcal{P}^Q_M) = \frac{\alpha + \beta + 2\gamma}{4}\]

We see that, among the plays $\mathcal{P}_{1, 2, M}$ and $\mathcal{P}^Q_{1, 2, M}$, the plays $\mathcal{P}^Q_{1, 2}$ are Pareto-Optimal. Thus the quantum version of the game allows players to achieve a pareto-optimal solution as a Nash Equilibrium, which is not possible in the classical case.

\section{Alternate Quantumization Protocol}
In certain scenarios, a special class of Classical Games can be quantumized in a different manner. This approach was followed by Meyer \cite{meyer1999quantum}, which we generalize in this section. We describe this class of Games as $\vb{G}(\mathcal{C}^{(k)}, \mathcal{M}, \pi)$. In this class of games, the players manipulate a common classical  system $\mathcal{C}$ which can be in be one of a finite set of classical states $\mathcal{C}^{(k)} = \{\mathcal{C}_1, \mathcal{C}_2, \ldots, \mathcal{C}_k\}$, and the players can select from a set of \emph{moves} $\mathcal{M} = \{M_1, M_2, \ldots, \}$ which transform the system from one state to another, i.e, 
\[M_i (\mathcal{C}_j) = \mathcal{C}_{ij}\qquad \text{where},\, \mathcal{C}_{ij}\in \mathcal{C}^{(k)}\]
Here, $\mathcal{C}_{ij}$ is called the Transition Matrix of the game. Depending on the structure of the game, a player can play a move multiple times. 

In these games, only the initial state of the system is known to the players, and the running state remains hidden as the game progresses. The final state is revealed at the end of the game only after all of the players have played their moves. The payoff of each player depends on this final state. 

For example, in the Penny-Flip game, the common Classical System is a penny with $\mathcal{C}^{(2)}$ = \{Heads (H), Tails (T)\}, the players moves are $\mathcal{M}$ = \{Flip (F), No-Flip (N)\}, and the payoff of the players depends on the final state of the penny being H or T, with $\pi_Q(H) = -\pi_Q(T) = \pi_C(T) = -\pi_C(H) = 1$. 

As an alternate to the EWL protocol, this class of games can also be quantumized as follows - 
\begin{enumerate}
    \item The Classical System $\mathcal{C}^{(k)}$ is replaced by a quantum system $\mathcal{Q}$ with
    \[\dim(\mathcal{Q}) = k\]
    which has orthonormal basis states $\{\ket{\mathcal{C}_1}, \ket{\mathcal{C}_2}, \ldots, \ket{\mathcal{C}_k}\}$
    \item The set of classical moves $\mathcal{M}$ is now replaced by the set of \emph{all} Unitary operators $U$ which can act on $\mathcal{Q}$. This set also contains the operators $U_{M_i}$ such that, 
    \[U_{M_i}\ket{\mathcal{C}_j} = \ket{\mathcal{C}_{ij}}\]
    These $U_{M_i}$'s are the so-called ``classical strategies'' in this quantum game. However, players may use quantum strategies or quantum moves other than these classical moves as well. A play of this quantum game $\mathcal{P}^Q$ consists of a sequence $\mathcal{U} = \ldots U_i\ldots U_2U_1$ of unitary operators played by the players. 
    \item $\mathcal{Q}$ is prepared in an initial state $\ket{\psi}$ known to the players. After all the players have played their respective Quantum Moves, the final state of $\mathcal{Q}$, $\ket{\psi_f} = \mathcal{U}\ket{\psi}$ is returned to a neutral judge. The judge measures $\mathcal{Q}$ in the $\{\ket{\mathcal{C}_i}\}$ basis. Depending on the measurement result, each player receives a payoff. The expected payoff of the $i$-th player is calculated as - 
    \begin{align*}
        \Bar{\pi}_i(\mathcal{U}) &= \sum_j p(\mathcal{C}_j) \cdot \pi_i(\mathcal{C}_j) \\
        &= \sum_j \bra{\psi_f}(\dyad{\mathcal{C}_i}) \ket{\psi_f}\cdot \pi_i(\mathcal{C}_j)\\
        &= \bra{\psi_f}\left(\sum_j \pi_i(\mathcal{C}_j)\dyad{\mathcal{C}_i}\right)\ket{\psi_f}\\
        &=\bra{\psi_f}\hat{\pi}_i\ket{\psi_f}\\        &=\bra{\psi}\mathcal{U}^{\dagger}\hat{\pi}_i\mathcal{U}\ket{\psi} \\
        &= \Tr [\dyad{\psi}\mathcal{U}^{\dagger}\hat{\pi}_i\mathcal{U}] \\
        &= \Tr[\rho\, \mathcal{U}^{\dagger}\hat{\pi}_i\mathcal{U}]\numberthis \label{eq_payoff_alt}
    \end{align*}
    where, 
    \[\hat{\pi}_i = \sum_j \pi_i(\mathcal{C}_j)\dyad{\mathcal{C}_i}\]
    is the Payoff Operator for player $i$, and 
    \[\rho = \dyad{\psi}\]
    is the initial state of $\mathcal{Q}$. 

    We can see that eq.~\ref{eq_payoff_alt} is the same as eq.~\ref{eq_payoff_main}, and thus the formula using Payoff Operators also holds in this case. This protocol can be easily generalized to the case where the starting state is a mixed state. The difference is in $\mathcal{U}$, where in eq. \ref{eq_payoff_main} the Unitary Operators are applied in '\emph{parallel}', whereas in eq.~\ref{eq_payoff_alt} they are applied in `\emph{sequence}'. 
\end{enumerate}

In Meyer's \cite{meyer1999quantum} quantumization of the Penny Flip game, $\mathcal{Q} = \mathcal{H}^{(2)}$ is a 2-dimensional quantum system, i.e, a qubit. The initial state is $\rho = \dyad{\vb{H}}$, and the sequence of operations the players use is \[\mathcal{U} = \textsf{U}_Q^{\star}\textsf{U}_C\textsf{U}_Q^{\star}\]
with $\textsf{U}_C \in \{\textsf{F}, \textsf{N}\}$ defined in eq.~\ref{eq_meyer_class}. The Payoff Operators are - 
\[\hat{\pi}_Q = \dyad{\vb{H}} - \dyad{\vb{T}} \qquad \hat{\pi}_C = \dyad{\vb{T}} - \dyad{\vb{H}}\]
In this language, 
\[\rho_f = \mathcal{U}\rho\mathcal{U}^{\dagger} = \dyad{\vb{H}}\]
So, the payoffs are - 
\[\Bar{\pi}_Q = \Tr[\hat{\pi}_Q\dyad{\vb{H}}] = 1 \qquad \Bar{\pi}_C = \Tr[\hat{\pi}_C\dyad{\vb{H}}] = -1 \]
As is evident from the payoffs, player Q always wins.

\section{Discussions and Conclusion}
In this exposition, we have reviewed the the foundational works of Meyer \cite{meyer1999quantum}, Eisert \etal \cite{eisert1999quantum, eisert2000quantum}, Marinatto \etal \cite{marinatto2000quantum} and Landsburg \cite{landsburg2004quantum, landsburg2011nash} in formulating Pure and Mixed strategy Quantum Games. We have generalized their works into two \emph{Quantumization Protocols}, the first of which is general and follows the approach of EWL, while the second is applicable to some special games and follows the approach of Meyer. In all of the cases, we have shown that the corresponding quantum games allow the players to achieve additional Quantum Nash Equilibria that Pareto-dominate the Classical Nash Equilibria. 

Existence of of Quantum Nash Equilibria are shown for a finite (and restricted) set of Quantum Strategies. This restriction allows us to use the mathematical techniques of finite games, and particularly Nash's theorem \cite{nash1950equilibrium} to prove the existence of Quantum Nash Equilibria. The case of the existence of Mixed Strategy Quantum Nash Equilibria for the entire set of Quantum Strategies ($U \in SU(2)$) requires more sophisticated techniques, since the strategy space then becomes infinite. To do this, we need to define a probability measure on $SU(2)$ or $SU(n)$, and start from there. Nevertheless, the framework that we have introduced is very general and can be used to perform such an analysis. Another avenue of investigation is what effect the initial state has on the achievable equilibria, as clearly, the final payoff depends on the initial state of the Quantum System.

In addition, the description of the dynamics of Quantum Games have been barely touched upon. This is because, the Payoffs of these games are calculated through Projective Quantum Measurements, which reveal the probabilities of the corresponding projectors and \emph{not} the probability `amplitudes' of the Quantum State. It is these amplitudes which are modified by Quantum Strategies and which control the evolution of the state. So, the dynamics of these Quantum Games have to be treated in a framework in which the players have some sort of \emph{incompleteness} in their knowledge of the strategies other players are using. Interestingly, there is actually very little work treating this dynamics, and most of the body of literature on Quantum Games deals with Static Quantum Games. We leave this exploration for future investigation.

\bibliographystyle{ieeetr}
\bibliography{reference.bib}

\end{document}